\documentclass{appolb}
\usepackage{epsfig}

\begin{document}
\title{Lattice QCD and Three-Body Exotic Systems in the static limit%
\thanks{talk presented at excited QCD 09, Zakopane, Poland}%
}
\author{
Marco Cardoso \address{CFTP, Instituto Superior T\'ecnico}
\and
Pedro Bicudo \address{CFTP, Instituto Superior T\'ecnico}
\and
Orlando Oliveira \address{CFC, Universidade de Coimbra}
}
\maketitle
\begin{abstract}
The static potentials for quark-antiquark-gluon and 3-gluon systems  are computed with lattice QCD methods.
For the quark-antiquark-gluon hybrid meson the static potential is obtained for different values of the angle between the quark-gluon and antiquark-gluon segments. The simulations support the formation of an adjoint string for small angles, while for large angles, the adjoint string is replaced by two fundamental strings connecting the gluon and the quarks.
For the 3-gluon  glueball, we discuss the corresponding Wilson loops and show that the gluons are connected by fundamental strings when the gluons are far apart.
\end{abstract}
  
\section{Introduction}

We explore, in Lattice QCD, the static potential of three-body systems with gluon(s)
using Wilson loops, namely the quark-antiquark-gluon hybrid and the three gluons glueball.
The interest in this systems is increasing because of the future experiments BESIII at IHEP in Beijin, GLUEX at 
JLab and PANDA at GSI in Darmstadt, dedicated to study the mass range of 
charmonium, with a focus in its plausible excitations and in hybrid and
glueball production.  The three-gluon glueballs are also relevant to the odderon. 
Thus several models are being developed and the static potential should, at least, provide one important component of the 
full interaction.
The first Lattice studies of the gluon interactions were performed by Michael \cite{Michael:1985ne,Campbell:1985kp}
and Bali extended them to other SU(3) representations \cite{Bali:2000un}.
Okiharu and colleagues \cite{Okiharu:2004ve,Okiharu:2004wy} studied tetraquarks and pentaquarks on the lattice.
Bicudo, Cardoso and Oliveira studied the hybrid quark-antiquark-gluon static potential,
\cite{Bicudo:2007xp,Cardoso:2007dc}.

\begin{figure}
\centering
\includegraphics[width=0.9\textwidth]{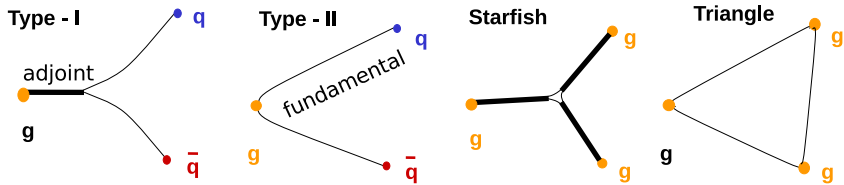}
\caption{The two models of confinement: type I and type II.}
\label{type1type2}
\end{figure}

We compare different models of confinement, namely the Casimir Scaling, type I and type II superconducor models.
In the type I model, adjoint strings are formed
while in the type II model, the fundamental strings repulse and adjoint strings are not formed.
This comes from the analogy with the two types of condensed matter superconductors (see Fig. \ref{type1type2}).

In the case of the hybrid meson type II models means a fundamental string linking the quark and the antiquark to the gluon, while type I  models require an adjoint string. For the three-gluon glueball, the type I I means three fundamental string linking the gluons in a triangle shape, while for the type I the three adjoint strings fuse at a common point giving rise to a starfish geometry (see Fig. \ref{type1type2}). 

\section{Wilson Loops}

To measure the static potentials in lattice QCD, we construct an operator
- the wilson loop - which corresponds to our system. The mean value of this
operator could be expanded in euclidean time as
\begin{equation}
	\langle W(t) \rangle = \sum_n C_n e^{ - V_n t }
\end{equation}
with $V_0$ being the static potential of the system. 

For the case of a meson this operator is simply given by a closed loop of lattice links
\begin{eqnarray}
	W_{q\bar{q}}(R,T) = Tr[ &
	U_{\mu}(0,0) \ldots U_{\mu}(R-1,0)
	U_{4}(R,0) \ldots U_{4}(R,T-1)
	\\ \nonumber
	& U_{\mu}(R-1,T)^{\dagger} \ldots U_{\mu}(0,T)^{\dagger}
	U_{4}(0,T-1)^{\dagger} \ldots U_{4}(0,0)^{\dagger}
	]
\end{eqnarray}

In the cases of hybrid mesons and glueballs, we have lines corresponding not only
to quarks and antiquarks, but also to gluons. Since the gluons are in the adjoint representation of $SU(3)$
the gluonic lines correspond to adjoint paths in the lattice $\tilde{U}^{ab} = \frac{1}{2} Tr[ \lambda^a U \lambda^b U^{\dagger} ] $. This operators could be simplified by using the Fierz relation:
\begin{equation}
	\sum_a \lambda^a_{ij} \lambda^{a}_{kl} =
	 2 \delta_{il} \delta_{jk} - \frac{2}{3} \delta_{ij} \delta_{kl}
	\label{fierz}
\end{equation}

\begin{figure}
\centering
\includegraphics[width=0.9\textwidth]{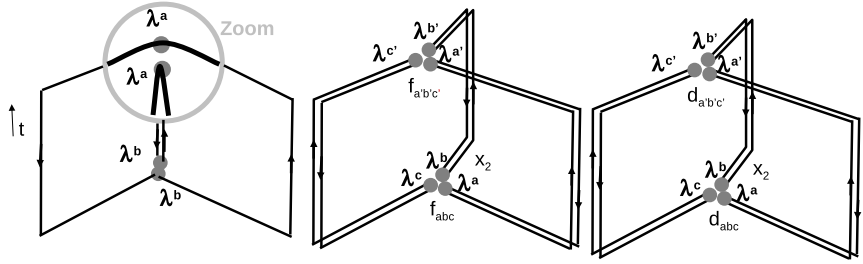}
\caption{Left: Quark-antiquark-gluon Wilson loop. Center and Right: Antisymmetric and Symmetric Three-gluons
Wilson loop. }
\label{loops}
\end{figure}
For the results int this work, we used 141 $24^3\times 48$ $SU(3)$ lattice configurations with $\beta=6.2$, generated
with MILC code \cite{MILC}.

\subsection{quark antiquark gluon Wilson Loop}

In the case of the quark-antiquark-gluon the wilson loop is given by \cite{Bicudo:2007xp,Cardoso:2007dc} (Fig. \ref{loops}):
\begin{eqnarray}
	W_{q\bar{q}g}=
	%
	\mbox{Tr} [
	&&
	U_4 (x,0) \ldots U_4 (0,t-1)  ~ \lambda^a ~
  	U^\dagger_4 (0,t-1) \ldots U^\dagger_4 (0,0)  ~  \lambda^b ~ ]
	\nonumber \\
	\mbox{Tr}
	[
	&&
	\lambda^b ~
	L_{1}(0) ~ U_{4}( \mathbf{x}_1 , 0 ) \ldots U_{4}( \mathbf{x}_1 , t - 1 ) ~ L_{1}^{\dagger}(t)
   	\nonumber \\
  	&&
	\lambda^a ~
   	L_{2}(t) ~ U_{4}^{\dagger}( \mathbf{x}_2 , t - 1 ) \ldots U_{4}^{\dagger}( \mathbf{x}_2 , 0 ) ~  L_{2}^{\dagger}(0)
	 ~ ] ~ .
\end{eqnarray}
where $L_i(t)$ is a spatial path that links the origin ( where the gluon is ) to the position $\mathbf{x}_i$ in the time
slice $t$. Using (\ref{fierz}), this operator becomes $W_{q\bar{q}g} \propto W_1 W_2 - \frac{1}{3} W_3$ where $W_1$, $W_2$ and $W_3$ are the three fundamental ( i. e. mesonic ) wilson loops.

\subsection{Three gluons Wilson Loop}

For the three gluons there are two possible color wavefunctions, corresponding to opposite charge conjugation
properties. One is antisymmetric and the other symmetric for the permutation of two gluons. 
For this case, the wilson loop operators have the form ( \cite{Cardoso:2008sb} )
\begin{equation}
	W_{3g}^{A/S} = T_{abc} T_{a'b'c'} \tilde{P_1}^{aa'} \tilde{P_2}^{bb'} \tilde{P_3}^{cc'}
\end{equation}
where $T_{abc} = f_{abc}$ for the antissymetric case and $T_{abc} = d_{abc}$ for the symmetric, and
\begin{equation}
	P_i = L_{i}(0) ~ U_{4}( \mathbf{x}_i, 0 ) \ldots U_{4}( \mathbf{x_i}, t - 1 ) ~ L_{i}^{\dagger}(t).
\end{equation}

The two three-gluon Wilson loops, as a function of fundamental paths are given by
\begin{equation}
\label{antisymmetricoperator}
	W_{3g}^{A} \propto  \mbox{Re} \big\{
		\mbox{Tr}[ P_1 P_2^\dagger ] \mbox{\mbox{Tr}}[ P_2 P_3^\dagger ] \mbox{Tr}[ P_3 P_1^\dagger ] 
		\big\}
		- \mbox{Re} \mbox{Tr}[ P_1 P_2^\dagger P_3 P_1^\dagger P_2 P_3^\dagger ]
\end{equation}
and
\begin{eqnarray}
	W_{3g}^S &\propto& \mbox{Re} \mbox{Tr}[ P_1 P_2^\dagger P_3 P_1^\dagger P_2 P_3^\dagger ] +
	\mbox{Re} \big\{ \mbox{Tr}[ P_1^\dagger P_2 ] \mbox{Tr}[ P_2^\dagger P_3 ] \mbox{Tr}[ P_3^\dagger P_2 ] \big\}
\nonumber \\ &&
	- \frac{2}{3} \mbox{Tr}[ P_1 P_2^\dagger ] \mbox{Tr}[ P_1^\dagger P_2 ]
	- \frac{2}{3} \mbox{Tr}[ P_2 P_3^\dagger ] \mbox{Tr}[ P_2^\dagger P_3 ]
	- \frac{2}{3} \mbox{Tr}[ P_3 P_1^\dagger ] \mbox{Tr}[ P_3^\dagger P_1 ]
\nonumber \\ &&
	+ \frac{4}{3} \ .
\label{symmetricoperator}
\end{eqnarray}

\section{Static Potential Results}

\subsection{Hybrid Meson}

For the hybrid meson, we study the static potential of the system as a function of the quark-gluon
distance ( $r_1$ ), the antiquark-gluon distance ( $r_2$ ), and the angle $\theta$ between the quark-gluon and
antiquark-gluon segments. By using off-axis directions on the lattice we can, not only compute the potential for
$\theta = 0^{\circ}, ~ 90^{\circ}$ and $180^{\circ}$, but also for $\theta = 45^{\circ}, ~ 60^{\circ}, ~ 120^{\circ}$
and $135^{\circ}$.

\begin{figure}
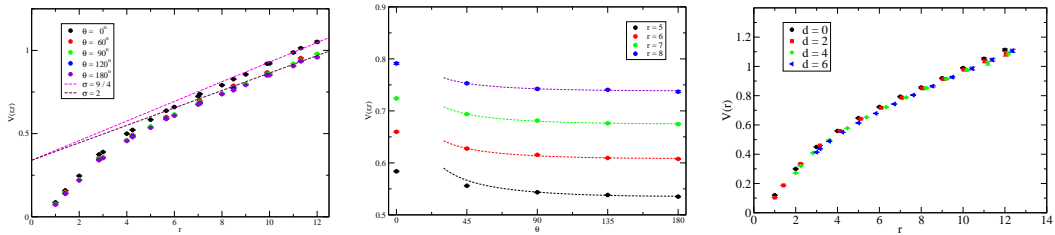

\begin{picture}(350,100)(0,0)
\put(-20,0){
\includegraphics[width=0.34\textwidth]{VRRsigma.eps}
}
\put(115,0) {
\includegraphics[width=0.34\textwidth]{qqgcoul.eps}
}
\put(250,0) {
\includegraphics[width=0.35\textwidth]{Upot.eps}
}
\end{picture}
\caption{Left: Results for the V(r,r) on the hybrid meson for various angles and comparision with two different string tensions. Middle: Results for V(r,r) as a function of $\theta$ for different values of $r$
Right: Results for the U geometry as function of $r_1 = r_2 = r$ for various values of quark antiquark distance.
}
\label{qqgpic}
\end{figure}

First, we study the results for the hybrid meson for the special case $r_1 = r_2 = r$, for the different values
of $\theta$. As can be seen in Fig. \ref{qqgpic}, for large $r$ the potentials are given by $V(r) \sim \sigma' r$.
We can also see that for all angles we have $\sigma' \simeq 2 \sigma$, except for $\theta = 0^{\circ}$, in which
case we have $\sigma' = \frac{9}{4} \sigma$, which is the value predicted by Casimir Scaling. This means that for
$\theta = 0^{\circ}$ an adjoint string is formed while for the other angles we essentialy have two independent fundamental strings.

Also in Fig. \ref{qqgpic}, we see the results for the potential with $r_1 = r_2 = r$
as a function of $\theta$ for various values of $r$.
This results are fitted to a coulomb like potential, which in this case changes only with $\theta$ in the
type II superconductor model. This model fits well the resuslts for large $\theta$ and $r$.

With this results, we can see the existence of two regimes - one with adjoint string formation ( for $\theta = 0$ )
and the other with fundamental strings. To better understand the transition between the two,
we are studying a geometry with a shape of a $U$, given in Fig. \ref{qqgpic}.

\subsection{Three gluons}

For the case of the three-gluon glueball, we study two geometries. One, in which the three gluons
are in three perpedincular axis and at the same distance of the origin, forming an equilateral triangle, and another
in which one of the three gluons is at the origin and the other two are in the axis and at the same distance of the
other gluon, drawing a rect isosceles triangle.

\begin{figure}
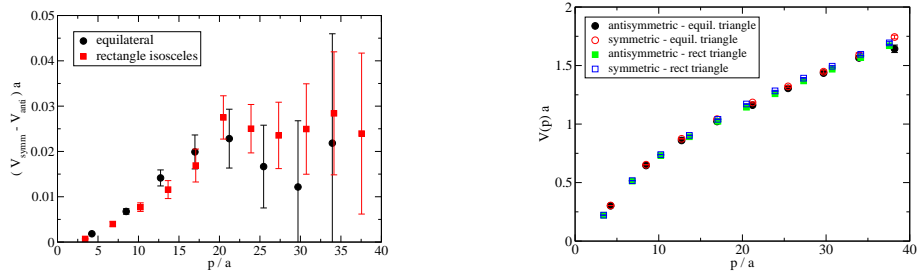

\begin{picture}(350,120)(0,0)
\put(20,0){
\includegraphics[width=0.4\textwidth]{Vdiff5.eps}
}
\put(220,0) {
\includegraphics[width=0.4\textwidth]{pot3g.eps}
}
\end{picture}
\caption{
Left: Results for the difference of the three gluon static potential in the two color arrangements
Right: Results for the three gluon potential as a function of the perimeter of the triangle formed by the three quarks.
}
\label{pot3g}
\end{figure}

For this geometries we study the difference between the two potentials and the two potentials separately.
Both results are shown on Fig. \ref{pot3g}.

For $V_{symm} - V_{anti}$ we see that the difference is systematicaly positive and rising linearly with the perimeter
of the triangle, with a fitted string tension $\sigma_{diff} = 0.04 \sigma$.

By fitting $V_{anti}$ and $V_{symm}$ to a potential of the form
$ V = C_0 - \alpha \sum_{i<j} \frac{1}{r_{ij}} + \sigma' p $, where p is the perimeter of the triangle, we get
$\sigma' = \sigma$, showing that the geometry of the strings is the triangle geometry and not the starfish geometry, which
would have given $\sigma' = \frac{9}{4\sqrt{3}} \sigma$ for the equilateral triangle, and
$\sigma' = \frac{9(1+\sqrt{3})}{8(1+\sqrt{2})} \sigma$ for the rect isosceles triangle.

\section{Conclusions}

From both the results we see that the confinement is essentialy described by a type II superconductor model,
except when the fundamental strings overlap, in which case adjoint strings are formed. This result is
important for constituent quark and gluon models. In the three-gluon glueball this type-II superconductor behaviour
manifests itself by the triangular string geometry. However this picture does not account for the formation of
adjoint strings and does not explain the little systematic difference between the static potential of 
the two color wavefunctions.

In order to understand better the formation of the adjoint string and the validity of the type II superconductor model, we
are computing the static distributions of the chromoeletric field in both systems.

\end{document}